# Smectic Pair Density Wave Order in EuRbFe$_4$As$_4$

He Zhao[1], Raymond Blackwell[1], Morgan Thinel[3,4],
Taketo Handa[4], Shigeyuki Ishida[2], Xiaoyang Zhu[4], Akira Iyo[2], Hiroshi Eisaki[2], Abhay N. Pasupathy[1,3*], Kazuhiro Fujita[1*]

[1]Condensed Matter Physics and Materials Science Department, Brookhaven National Laboratory, Upton, NY, 11973 USA
[2]Research Institute for Advanced Electronics and Photonics, National Institute of Advanced Industrial Science and Technology, Tsukuba 305-8568, Japan
[3]Department of Physics, Columbia University, New York, NY 10027 USA
[4]Department of Chemistry, Columbia University, New York, NY 10027 USA

**The pair density wave (PDW) is a novel superconducting state in which Cooper pairs carry center-of-mass momentum in equilibrium, leading to the breaking of translational symmetry[1,2,3,4]. Experimental evidence for such a state exists in high magnetic field[5,6,7,8] and in some materials that feature density wave orders that explicitly break translational symmetry[9,10,11,12,13]. However, evidence for a zero-field PDW state that exists independent of other spatially ordered states has so far been elusive. Here, we show that such a state exists in the iron pnictide superconductor EuRbFe$_4$As$_4$ (Eu-1144), a material that features coexisting superconductivity ($T_c$ ~ 37 K) and magnetism ($T_m$ ~ 15 K)[14,15]. We show from the Spectroscopic Imaging Scanning Tunneling Microscopy (SI-STM) measurements that the superconducting gap at low temperature has long-range, unidirectional spatial modulations with an incommensurate period of ~8 unit cells. Upon raising the temperature above $T_m$, the modulated superconductor disappears, but a uniform superconducting gap survives to $T_c$. When an external magnetic field is applied, gap modulations disappear inside the vortex halo. The SI-STM and bulk measurements show the absence of other density wave orders, showing that the PDW state is a primary, zero-field superconducting state in this compound. Both four-fold rotational symmetry and translation symmetry are recovered above $T_m$, indicating that the PDW is a smectic order.**

1.      The application of a large magnetic field results in a Zeeman splitting in the energies of spin up and down electrons. This spin-splitting is deleterious to uniform superconducting order based on paired singlets. Fulde, Ferrel, Larkin and Ovchinnikov (FFLO) predicted that an alternate superconducting state where the Cooper pairs possess a non-zero center of mass momentum can be stabilized in this situation[16,17]. The presence of a non-zero center of mass momentum leads to a superconducting order parameter that is periodically modulated in real space ($\Delta_Q(\mathbf{r})$). Evidence for this elusive FFLO state has been seen in heavy fermion materials[18,19,20] and organic metal compounds[8,21,22,23,] in high applied magnetic field via thermodynamic and nuclear magnetic resonance measurements. However, direct visualization of $\Delta_Q(\mathbf{r})$ from the FFLO mechanism has so far not been achieved.

2.      More recently, it has been theoretically realized that there could be several situations beyond the original FFLO proposals in which spatially modulated superconductivity exists[1]. A simple way this can occur is if a material features both the uniform superconducting and (charge or spin) density wave orders. Since both the superconductivity and density wave order feature the same charge and spin degrees of freedom, they are inevitably coupled, leading to modulated superconducting order parameters in real space. This modulated superconductivity is a secondary order – a consequence of the coupling between the primary orders of uniform superconductivity and charge or spin density waves. Alternatively, it is possible that a spatially modulated superconducting state is stabilized as a primary superconducting instability without an external magnetic field. This state is termed a pair density wave (PDW).

3.      In the past few years, spatially modulated superconducting gaps that ride on a background of a much larger uniform gap have been directly visualized by STM spectroscopy in cuprate superconductors[5,9] Kagome compounds[11], and two-dimensional chalcogenides[10]. In all of these materials, robust density wave orders exist independent of superconductivity[24]. Arguments have been advanced that the superconducting modulations are a primary PDW order in the cuprates and the Kagome compounds[5,9]. However, the experimental situation remains unclear. A direct visualization of superconducting modulations that can be clearly tied to a primary PDW instability is therefore valuable to advance our understanding of this state of matter in quantum materials [1].

4. Here we describe key signatures of the PDW state that are expected to appear in STM data. First, we consider a PDW, which is defined as a spatially modulated superconducting order parameter $\Delta_{\mathbf{Q}}(\mathbf{r}) = \Delta_1[e^{i\mathbf{Q}\cdot\mathbf{r}} + e^{-i\mathbf{Q}\cdot\mathbf{r}}]$ (LO state), where $\Delta_1$ is a positive real number. STM spectroscopy measures the local density of states as a function of energy, $N(\mathbf{r}, E)$. The most obvious and immediate signature of the PDW is that $N(\mathbf{r}, E)$ should detect a local gap, $|\Delta(\mathbf{r})|$, which is identical to the local order parameter, $|\Delta_{\mathbf{Q}}(\mathbf{r})|$, when the coherence length is short compared to the PDW wavelength, and modulates at $\mathbf{Q}$. If such a PDW coexists with spatially uniform superconducting order with energy gap $\Delta_0$, it is expected that $N(\mathbf{r}, E)$ displays strong modulations around $E \sim \Delta_0$. Therefore, the second unique signature of a PDW with wavevector $\mathbf{Q}$ that coexists with uniform superconductivity would be a presence of $N(\mathbf{r}, E \sim \Delta_0)$ modulations at $\mathbf{Q}$. In reality, the superconducting coherence length[18] in Eu-1144 is comparable to the PDW wavelength we measure. In this situation, the gap measured in $N(\mathbf{r}, E)$ cannot simply be assigned to the local superconducting order parameter, but is instead a complicated non-local quantity[25]. However, the two signatures described above – local variations of the gap measured in $N(\mathbf{r}, E)$ and variations of $N(\mathbf{r}, E \sim \Delta_0)$ – continue to be good evidence for the presence of PDW order.

**Topographic characterization**

5. EuRbFe$_4$As$_4$ (Eu-1144) consists of alternately stacked "122" layers of EuFe$_2$As$_2$ and RbFe$_2$As$_2$ as shown in Fig. 1a. This compound exhibits both high $T_c$ superconductivity ($T_c \sim 37$ K) and helical magnetic order ($T_m \sim 15$ K, Fig. 1b). The magnetic order arises from in-plane ferromagnetism in the Eu$^{2+}$ plane, whose ordering direction rotates by 90 degrees between adjacent Eu$^{2+}$ planes in a chiral fashion[15]. Single crystals of Eu-1144 are synthesized by the RbAs-flux method[26] (Extended data 1) and are cleaved at $T \ll 70$ K prior to insertion into the STM head (see methods).

6. Like RbFe$_2$As$_2$ [28], cleaving the sample is expected to result in fracture in the Rb plane and a loss of half of the atoms in the Rb plane to maintain charge neutrality. Fig. 1d is a representative topograph $T(\mathbf{r})$ of the Eu-1144 surface within a 5 nm × 5 nm field-of view (FOV). Individual Rb atoms are clearly resolved, and the observed lattice constants are $\sqrt{2} \times \sqrt{2}$ times larger than the uncleaved Rb plane, as expected[28,29]. Note that the Fe-Fe directions are the same as those for Rb-Rb of the $\sqrt{2} \times \sqrt{2}$ lattice, as indicated by the arrows in Fig. 1d, while the lattice constant of the Fe-Fe lattice is one-half of the Rb-($\sqrt{2} \times \sqrt{2}$) lattice. Fig. 1e shows a large, 110 nm × 110 nm FOV $T(\mathbf{r})$ of the Rb surface. Missing atoms (dark spots) and impurities (bright spots) are observed on this large length scale. Fig. 1f is the Fourier transform (FT) $T(\mathbf{q})$ of the $T(\mathbf{r})$ shown in Fig. 1e. An absence of any superlattice peaks in $T(\mathbf{q})$ indicates the absence of surface reconstructions or other density wave orders (additional topographic images are shown in the Extended data Fig. 4). In Fig. 1g, we show the spatially averaged local density of states spectra $N(E)$ at different temperatures from $T = 8$ K to 35 K. At $T = 8$ K, particle-hole symmetric coherence peaks are clearly observed with a gap $\Delta_0$ = 6 mV. With increasing temperature, the gap weakens and eventually closes at the bulk $T_c$. No sudden changes are detected in $\Delta_0$ at the magnetic transition temperature $T_m$. These results establish that a well-defined, uniform superconducting gap $\Delta_0$ is present in STM spectroscopy at all temperatures below bulk $T_c$.

## Unidirectionally-modulated superconducting gap

7. We next proceed to measure the local spectroscopic properties of Eu-1144. Shown in Fig. 2a is a map of the local gap values $|\Delta(\mathbf{r})|$ measured within a 30nm × 30nm FOV. Remarkably, a unidirectionally modulated $|\Delta(\mathbf{r})|$ along the Fe-Fe direction is clearly resolved. The spatial variation of $|\Delta(\mathbf{r})|$ is well described by the form $|\Delta(\mathbf{r})| = |\Delta_0| + |\Delta_1|\cos(\mathbf{Q}_{PDW} \cdot \mathbf{r})$, with $|\Delta_0| = 6$ mV and $|\Delta_1| = 0.6$ mV. We see clearly from this image that the unidirectional modulations in $|\Delta(\mathbf{r})|$ persist across the image with a uniform period. The FT of Fig. 2a is shown in Fig. 2b, that exhibits clear $\mathbf{Q}_{PDW}$ peaks (circled) corresponding to the modulation of $|\Delta(\mathbf{r})|$ seen in Fig. 2a. A line cut of the FT in Fig. 2b through the peak is displayed in Fig. 2c, showing that the width of the peak is essentially a single pixel over this FOV. The observed incommensurate wavelength is about eight unit-cells. Direct inspection of the modulations in Fig. 2a as well as the sharpness of the $\mathbf{Q}_{\text{PDW}}$ peak in Fig. 2b indicates that the coherence length of the $|\Delta(\mathbf{r})|$ modulations is larger than 30 nm (SI section 1). Our observation of the spatially modulated energy gaps is our primary result and indicates the presence of electron pairs with a finite center-of-mass momentum.

8. The spatial modulations of superconductivity are also directly observed in $N(\mathbf{r}, E \sim \Delta_0)$, as shown in Fig. 2d (Extended data Fig. 3). A set of $N(\mathbf{r}, E)$ spectra obtained along the dashed line in Fig. 2a are displayed in Fig. 2e. These spectra show particle-hole symmetric gap modulations. The Fig. 2e inset shows the spatial oscillations of $|\Delta(\mathbf{r})|$ clearly near the coherence peaks. In order to examine the energy dependence of these modulations, we plot in Fig. 2f the energy dependence of the $N(\mathbf{q}, E)$ maps in the direction of the ordering vector (SI section 3 and Extended data Fig. 5). We can see clearly that sharp features are observed at $\mathbf{Q}_{\text{PDW}}$ only around the superconducting gap energies and are not observed away from these energies (SI section 2). In addition, these features do not change with energy.

9. The absence of modulations at high energies provides evidence against the presence of other ordered states such as charge or spin density waves, at least within the resolution of SI-STM (SI sections 3 and 5). Further, no density wave orders have been reported to date by elastic scattering measurements in Eu-1144[26]. To investigate even further the possibility of a primary charge density wave order, we performed Raman spectroscopy measurements of the phonon modes of Eu-1144 as a function of temperature. Analysis of the wavenumbers of observed phonon modes (see extended data Fig. 6) shows no evidence for the softening near the magnetic transition temperature. No additional modes that might arise due to zone folding are observed below the magnetic transition. The Raman measurements therefore provide strong additional evidence for the absence of a primary density wave order in Eu-1144. Since the observed local density of states modulations near the gap energies are neither due to surface reconstructions, quantum interference of the quasiparticle in a superconductor[30] nor due to alternative orders[31], the only remaining possibility is that these modulations indicate a primary PDW superconducting instability in this compound.

## Temperature-dependent pair density wave

10. We next study the temperature dependence of the phenomena observed in Figure 2. Shown in Figs. 3a-e are a sequence of $N(\mathbf{r}, E \sim \Delta_0)$ maps measured at $T = 4.5$ K, 10 K, 15 K and 18 K respectively (Extended data Fig. 2). It is obvious that with increasing temperature the modulations become weaker and disappear around the magnetic transition temperature $T_m$. We extract the intensity of the modulations at $\mathbf{Q}_{PDW}$ in FTs of Figs. 3a-d at each temperature (Figs. 3e-h) with all data obtained under identical setpoint conditions. The result is plotted in Fig. 3i, in which values are normalized by the one at 4.5 K. Our conclusion is that the modulated part of the superconducting gap $|\Delta_1|$ has roughly a mean field temperature dependence and disappears precisely when magnetism of the Eu plane is lost. Line cuts of the FT of $N(\mathbf{r}, E)$ at 4.5 K (below $T_m$) and 16 K (above $T_m$) from a second set of independent measurements on a different sample are shown in Figs. 3j,k, clearly showing the complete absence of $\mathbf{Q}_{PDW}$ above $T_m$.

## Suppression of pair density wave inside magnetic vortices

11. We now turn to the magnetic field dependence of the observed phenomena. Shown in Fig. 4a is $N(\mathbf{r}, E \sim \Delta_0)$ measured in a field of 9 Tesla applied along the *c*-axis. Superconducting vortices that exhibit reduced $N(\mathbf{r}, E \sim \Delta_0)$ are clearly seen forming a quasi-triangular vortex lattice. The unidirectional modulations in $N(\mathbf{r}, E \sim \Delta_0)$ are also clearly seen between the vortices. Fig. 4b is a FT of the entirety of Fig. 4a, which shows the peak at $\mathbf{Q}_{PDW}$. To examine the difference between the interior and exterior of the vortex, we apply a mask to select only the vortex areas, as shown in Fig. 4c. The FT of this masked area shows no peak at $\mathbf{Q}_{PDW}$ (Fig. 4d). To make sure that our masking procedure is not producing this result as an artefact, we simply shift the mask out of the vortices and apply it to Fig. 4a, resulting in the masked image in Fig. 4e. The FT of Fig. 4e is shown in Fig. 4f, clearly showing the peak at $\mathbf{Q}_{PDW}$. This shows that our conclusion is robust to the masking procedure applied, i.e., there are no density of states modulations inside the superconducting vortices. This is consistent with the expectation for a PDW coexisting with a uniform superconductor – the suppression of the dominant uniform superconducting gap $\Delta_0$ implies that no modulations should be observed in $N(\mathbf{r}, E \sim \Delta_0)$. Analysis of the field dependence for the PDW modulations indicates that the strength of the PDW outside the vortex halo does not depend on the applied field, at least up to 9 T (SI section 6).

12. The PDW that we observed is clearly unidirectional, implying that four-fold rotational ($C_4$) symmetry is broken in the PDW state. There are two trivial possibilities to explain this phenomenon. First, the magnetic moment of the $Eu^{2+}$ points in plane at zero field and breaks $C_4$ symmetry. Second, we speculate that strain in the crystal may exist that would break $C_4$ symmetry. The first possibility can be ruled out by the magnetic field dependence seen in Figure 4. Given that the out-of-plane saturation field of the $Eu^{2+}$ magnetic moment is about 0.4 Tesla[32], the unidirectionality of the PDW outside the vortex cores at 9T is not related to the magnetic moment direction. To examine the possibility of strain effects, we perform Bogoliubov quasiparticle interference measurements below and above $T_m$ (see extended data Fig. 7and SI section 4). These measurements show within our resolution that the electronic structure is $C_4$ symmetric above $T_m$, which is also consistent with the tetragonal structure of Eu-1144 from other measurements. Therefore, we conclude that the unidirectional order of the PDW is a spontaneously broken symmetry, i.e., the PDW is a smectic order that breaks both rotational and translational symmetries in this material.

13. Finally, we speculate the mechanism of PDW formation together with the uniform superconductivity in Eu-1144. One possibility is an exchange splitting of the Fermi surfaces due to coupling between the Fe-derived $d$ orbitals and the Eu $f$ electrons[33]. The Fermi surface of Eu-1144 features multiple hole pockets at the $\Gamma$ point and electron pockets at the M point, similar to other pnictide superconductors in this family[34,35,36,37]. In general, we expect distinct exchange couplings at the $\Gamma$ and M points, due to the very different orbital character of the bands. In this simple picture, we postulate that the electron pocket(s) (at the M points) along one of the principal axes host PDW order with a Q determined by the spin-splitting, while the hole pocket(s) near the $\Gamma$ point, which are relatively unaffected, support uniform superconductivity. An opposite situation, in which a role of the hole and electron pockets are switched in the discussion above would also be possible. In reality, scattering between pockets necessarily implies a PDW component on both pockets, but the schematic picture remains largely intact. In such a scenario, one might expect to see some temperature dependence of the **Q** vector due to a decrease in the exchange splitting with temperature. This is not observed in the experiment so far. Detailed future calculations can shed light on the expected temperature dependence of **Q**, and help verify or rule out this mechanism of the PDW formation.

14. Regardless of whether an exchange splitting explanation suffices to explain the presence of the PDW, or whether a more strong-coupling starting point is required, the data show the presence of a strong interaction between adjacent magnetic and superconducting planes in Eu-1144. This also points us towards a way for engineering a PDW order from the ground up, by coupling ultrathin magnetic materials and superconductors[38]. The availability of a wide variety of van der Waals materials and their use to create artificial superconductor-magnet heterojunctions points to a promising new direction to engineer the PDW state.

**References**

1. Agterberg, D. F. *et al.* The Physics of Pair-Density Waves: Cuprate Superconductors and Beyond. *Annu. Rev. Condens. Matter Phys.* **11**, 231–270 (2020).
2. Berg, E. *et al.* Dynamical Layer Decoupling in a Stripe-Ordered High-$T_c$ Superconductor. *Phys. Rev. Lett.* **99**, 127003 (2007).
3. Casalbuoni, R. & Nardulli, G. Inhomogeneous superconductivity in condensed matter and QCD. *Rev. Mod. Phys.* **76**, 263 (2004).
4. Lee, P. A. Amperean Pairing and the Pseudogap Phase of Cuprate Superconductors. *Phys. Rev. X.* **4**, 031017 (2014).
5. Edkins, S. D. *et al.* Magnetic field-induced pair density wave state in the cuprate vortex halo. *Science* **364**, 976–980 (2019).
6. Radovan, H. A. *et al.* Magnetic enhancement of superconductivity from electron spin domains. *Nature* **425**, 51–55 (2003).
7. A Gurevich. Iron-based superconductors at high magnetic fields. *Rep. Prog. Phys* **74**, 124501 (2011).
8. Bergk, B. *et al.* Magnetic torque evidence for the Fulde-Ferrell-Larkin-Ovchinnikov state in the layered organic superconductor $\kappa$-(BEDT-TTF)$_2$Cu(NCS)$_2$. *Phys. Rev. B* **83**,

064506 (2011).

9. Du, Z. *et al.* Imaging the energy gap modulations of the cuprate pair-density-wave state. *Nature* **580**, 65–70 (2020).

10. Liu, X., Chong, Y. X., Sharma, R. & Davis, J. C. S. Discovery of a Cooper-pair density wave state in a transition-metal dichalcogenide. *Science* **372**, 1447–1452 (2021).

11. Chen, H. *et al.* Roton pair density wave and unconventional strong-coupling superconductivity in a topological kagome metal. *Nature* **599**, 222–228 (2021).

12. Kinjo, K. *et al.* Superconducting spin smecticity evidencing the Fulde-Ferrell-Larkin-Ovchinnikov state in $Sr_2RuO_4$. *Science* **376**, 397–400 (2022).

13. Li, Q., Hücker, M., Gu, G. D., Tsvelik, A. M. & Tranquada, J. M. Two-dimensional superconducting fluctuations in stripe-ordered $La_{1.875}Ba_{0.125}CuO_4$. *Phys. Rev. Lett.* **99**, 067001 (2007).

14. Jiao, W.-H. *et al.* Iron-based magnetic superconductors $AEuFe_4As_4$ (A= Rb, Cs): natural superconductor-ferromagnet hybrids. *J. Phys. Condens. Matter* **34**, 21 (2022).

15. Iida, K. *et al.* Coexisting spin resonance and long-range magnetic order of Eu in $EuRbFe_4As_4$. *Phys. Rev. B* **100**.014506 (2019).

16. Fulde, P. & Ferrell, R. A. Superconductivity in a Strong Spin-Exchange Field. *Phys. Rev.* **135**, A550 (1964).

17. A. I. Larkin, Y. N. Ovchinnikov, Nonuniform state of superconductors. *Sov. Phys. JETP* **20**, 762–762 (1965).

18. Bianchi, A., Movshovich, R., Capan, C., Pagliuso, P. G. & Sarrao, J. L. Possible Fulde-Ferrell-Larkin-Ovchinnikov superconducting state in $CeCoIn_5$. *Phys. Rev. Lett.* **91**, 187004 (2003).

19. Kenzelmann, M. *et al.* Coupled superconducting and magnetic order in $CeCoIn_5$. *Science* **321**, 1652–1654 (2008).

20. Matsuda, Y. & Shimahara, H. Fulde–Ferrell–Larkin–Ovchinnikov State in Heavy Fermion Superconductors. *J. Phys. Soc. Jpn.* **76**. 051005 (2007).

21. Yonezawa, S. *et al.* Magnetic-Field Variations of the Pair-Breaking Effects of Superconductivity in $(TMTSF)_2ClO_4$. *J. Phys. Soc. Jpn.***77**.054712 (2008).

22. Shimahara, H. Theory of the Fulde–Ferrell–Larkin–Ovchinnikov State and Application to Quasi-Low-dimensional Organic Superconductors. *Springer Ser. Mater.*

*Sci.* **110**, 687–704 (2008).

23. Lortz, R. *et al.* Calorimetric evidence for a Fulde-Ferrell-Larkin-Ovchinnikov superconducting state in the layered organic superconductor $\kappa$-(BEDT-TTF)$_2$Cu(NCS)$_2$. *Phys. Rev. Lett.* **99**, 187002 (2007).

24. Himeda, A., Kato, T. & Ogata, M. Stripe States with Spatially Oscillating d-Wave Superconductivity in the Two-Dimensional $t – t' – J$ Model. *Phys. Rev. Lett.* **88**, 117001 (2002).

25. Baruch, S. & Orgad, D. Spectral signatures of modulated d-wave superconducting phases. *Phys. Rev. B* **77**, 174502 (2008).

26. Ishida, S. *et al.* Superconductivity-driven ferromagnetism and spin manipulation using vortices in the magnetic superconductor EuRbFe$_4$As$_4$. *Proc. Natl. Acad. Sci. U. S. A.* **118**, e2101101118 (2021).

27. Bao, J. K. *et al.* Single Crystal Growth and Study of the Ferromagnetic Superconductor RbEuFe$_4$As$_4$. *Cryst. Growth Des.* **18**, 3517–3523 (2018).

28. Liu, X. *et al.* Evidence of nematic order and nodal superconducting gap along [110] direction in RbFe$_2$As$_2$. *Nat. Commun.* **10**, 1039 (2019).

29. Liu, Y. *et al.* Superconductivity and ferromagnetism in hole-doped EuRbFe$_4$As$_4$. *Phys. Rev. B* **93**, 214503 (2016).

30. Fujita, K. *et al.* Simultaneous transitions in cuprate momentum-space topology and electronic symmetry breaking. *Science* **344**, 612–616 (2014).

31. Zhao, H. *et al.* Cascade of correlated electron states in the kagome superconductor CsV$_3$Sb$_5$. *Nature* **599**, 216–221 (2021).

32. Smylie, M. P. *et al.* Anisotropic superconductivity and magnetism in single-crystal EuRbFe$_4$As$_4$. *Phys. Rev. B* **98**, 104503 (2018).

33. Kim, T. K. *et al.* Electronic structure and coexistence of superconductivity with magnetism in EuRbFe$_4$As$_4$. *Phys. Rev. B* **103**, 174517 (2021).

34. Liu, W. *et al.* A new Majorana platform in an Fe-As bilayer superconductor. *Nat. Commun.* **11**, 5688 (2020).

35. Watson, M. D. *et al.* Probing the reconstructed Fermi surface of antiferromagnetic BaFe$_2$As$_2$ in one domain. *npj Quantum Mater.* **4**, 1–9 (2019).

36. Miao, H. *et al.* Observation of strong electron pairing on bands without Fermi

surfaces in LiFe$_{1-x}$Co$_x$As. *Nat. Commun.* **6**, 1–6 (2015).

37. Watson, M. D. *et al.* Three-dimensional electronic structure of the nematic and antiferromagnetic phases of NaFeAs from detwinned angle-resolved photoemission spectroscopy. *Phys. Rev. B* **97**, (2018).

38. Buzdin, A. I. Proximity effects in superconductor-ferromagnet heterostructures. *Rev. Mod. Phys.* **77**, 935 (2005).

**Figure 1. Characterization of Rb surface in EuRbFe$_4$As$_4$. a**, topographic three dimensional image of the Rb-terminated Eu-1144 surface encompassing a 1-unit-cell step. **b,c**, Top and side views of the crystal structure with the magnetic structure superimposed. **d**, Atomic resolution topograph showing individual Rb atoms. **e**, Large FOV topograph of Rb surface showing random Rb vacancies. **f**, Fourier transform of **e**, which shows an absence of any signatures of the surface reconstruction and density wave order. **g**, Spatially averaged differential conductance (d$I$/d$V$) spectra acquired over the area in **e** and its temperature evolution. Clear coherence peaks at $\pm\Delta$ indicate the superconducting gap size. The features at $\pm 23$ mV likely arise from electron-boson coupling. STM setup conditions: **d**, $V_{sample}$ = -10 mV, $I_{set}$ = 50 pA; **e,** $V_{sample}$ = 100 mV, $I_{set}$ = 40 pA; **g**, $V_{sample}$ = 30 mV, $I_{set}$ = 200 pA, $V_{exc}$ = 0.3 mV.

**Figure 2. Visualizing spatial modulations of the superconducting gap. a,** Superconducting gap map ($|\Delta(\mathbf{r})|$) showing unidirectional modulations. **b**, Fourier transform $|\Delta(\mathbf{q})|$ of a, showing a peak at $\mathbf{Q}_{PDW}$. The green crosses at $(0, \pm\frac{1}{4})$ and $(\pm\frac{1}{4}, 0)$ are along the Rb-($\sqrt{2}, \sqrt{2}$) square lattice. **c** Linecut of the Fourier transform in **b** through the $\mathbf{Q}_{PDW}$ peak, showing a single pixel width. **d**, $N(\mathbf{r}, E \sim \Delta_0)$ map near the average superconducting gap energy, $\Delta_0$ showing the same unidirectional modulations as in the gap map in **a. e**, $N(\mathbf{r}, E)$ spectra along the white dashed line in **d**. The inset is the intensity map around the coherence peak energy, clearly showing the modulation of the coherence peaks in energy. **f**, Map of line cuts of the Fourier transform of $N(\mathbf{r}, E)$ through $\mathbf{Q}_{PDW}$ as a function of energy. The $\mathbf{Q}_{PDW}$ peak is only observed near $E \sim \Delta_0$. STM setup conditions: $V_{set}$ = 12 mV, $I_{set}$ = 500 pA, $V_{exc}$ = 0.3 mV.

**Figure 3. Temperature dependence of PDW gap modulations. a–d**, Temperature-dependent $N(\mathbf{r}, E \sim \Delta_0)$ maps showing the gradual disappearance of the PDW gap modulation, which is completely suppressed above the magnetic transition temperature ($T_m \sim$ 15 K). All maps were taken in the identical region. **e–h**, The associated Fourier Transforms of **a-d**. **i**, A plot of the extracted intensities for the $\mathbf{Q}_{PDW}$ peak as a function of temperature. All the peak intensities are normalized by the one at 4.5K. **j,k,** Linecuts of the Fourier Transforms from **e** to **h** through $\mathbf{Q}_{PDW}$ as a function of energy for $T$=4.5 K and 16 K, respectively, for the second sample. The $\mathbf{Q}_{PDW}$ signals are completely suppressed at 16 K. STM setup conditions: $V_{set}$ = 6 mV, $I_{set}$ = 50 pA, $V_{exc}$ = 1 mV.

**Figure 4. Magnetic field dependence of the PDW gap modulations. a**, $N(\mathbf{r}, E \sim \Delta_0)$ map taken at 9 Tesla of the magnetic field showing a hexagonal vortex lattice. The vortex halos are marked by blue circles. The red circles are a uniform translation of the blue circles to regions outside the vortex halos. **b**, The Fourier transform of **a**. **c**, The same data as in **a** but with a mask applied for preserving only the vortex halo regions. **d**, Fourier transform of **c** showing the absence of the $\mathbf{Q}_{PDW}$ peak inside the vortex halo. **e** The same data as in **a** but with a mask applied for preserving only the regions in the red circles in a. **f.** Fourier transform of e showing the presence of the $\mathbf{Q}_{PDW}$ peak outside the vortex halo. STM setup conditions: $V_{set}$ = 6 mV, $I_{set}$ = 50 pA, $V_{exc}$ = 1 mV.

**Methods**

**Sample preparation.** To avoid a possible reconstruction of the surface for the cleaved sample, we cleaved EuRbFe$_4$As$_4$ single crystals at low temperature: we inserted into the STM head the uncleaved sample with a metallic cleaving rod that was glued on top of the sample and kept it for 30 minutes to reach 4.5 K. Then, we brought the sample back to the UHV preparation chamber, cleaved, and reinserted the cleaved one into the STM head. These procedures were performed in less than one minute. While we cannot exactly determine the temperature of the sample at the moment of cleaving, we estimate that it is well below 70 K (based on an observation of the peak temperature at the STM head after the sample insertion).

**STM experiments.** We studied two different pieces of the EuRbFe$_4$As$_4$ single crystals, both of which exhibited qualitatively the same phenomena as described in the main text. STM data were acquired using a customized Unisoku USM1300 microscope at different temperatures and magnetic fields along *c*-axis as denoted in the figure captions. Differential conductance (local density of states) measurements were performed using the standard lock-in technique with 2.5 KHz frequency and bias excitation as also detailed in the figure captions. We used electro-chemically etched tungsten tips for scanning that were prepared by applying a voltage pulse and/or fast scanning on the surface of the Au(111) single crystals (99.99%) prior to measurements on Eu-1144.

**Raman scattering experiments.** Raman scattering experiments were performed based on a home-built setup with a 633-nm HeNe laser and an inverted microscope. The laser is first sent through a linear polarizer, which fixes the polarization at the light source. The laser is then sent to a diffraction beamsplitter filter, sent through a half-wave plate that controls the incident polarization angle, directed to an inverted microscope objective (40x, numerical aperture = 0.6), and focused onto a sample mounted in a continuous-flow cryostat connected to a recirculating cooler system. The backscattered light from the sample is collected using the same objective and sent back through the same half-wave plate and diffraction beamsplitter filter. The backscattered signal is then sent through two volumetric Bragg filters to suppress the Rayleigh scattering. Parallel-polarized Raman signal was collected using a half-wave plate and fixed linear polarizer after the Bragg filters. The signal was recorded using a visible spectrometer with a high-resolution holographic grating and a liquid nitrogen-cooled charge-coupled device camera. Excitation power was 2.4 mW, and the Raman spectra shown are obtained with the incident polarization angle giving the largest Raman signal.

**Data availability**
All data that support the findings of this study are available from the corresponding author upon reasonable request. Source data are provided with this paper.

**Code availability**
The computer code used for data analysis is available upon request from the corresponding author.

**Acknowledgements** We thank Andreas Kreisel, Rafael Fernandes, Jörg Schmalian, Igor Mazin, Steven A. Kivelson, Yoichi Higashi, Yoichi Yanase, Hideo Aoki and Shin-ichi Uchida for fruitful discussions. Work at Brookhaven is supported by the Office of Basic Energy Sciences, Materials Sciences and Engineering Division, U.S. Department of Energy under Contract No. DE-SC0012704. Raman Spectroscopy measurements at Columbia University are supported by supported by the National Science Foundation MRSEC program in the Center for Precision-Assembled Quantum Materials under award number DMR-2011738 and by the Air Force Office of Scientific Research via grant FA9550-21-1- 0378. The work at AIST was supported by the Grant-in-Aid for Scientific Research on Innovative Areas "Quantum Liquid Crystals" (KAKENHI Grant No. JP19H05823) from JSPS of Japan.

**Author contributions** K.F. and A.N.P led the project.; H.Z. carried out SI-STM experiments with contribution from R.B. at Brookhaven National Laboratory; S.I., I. A. and H.E. synthesized and characterized the samples; M.T., T.H. and X.Y.Z carried out Raman spectroscopy measurements. H.Z. carried out analysis with contribution from R.B. K.F., A.N.P. and H.Z. wrote the manuscript. The manuscript reflects the contributions and ideas of all authors.

**Competing interests** The authors declare no competing interests.

**Additional information**

**Correspondence and requests for materials** should be addressed to Kazuhiro Fujita (kfujita@bnl.gov) and Abhay Pasupathy (apn2108@columbia.edu).

**Extended data Fig. 1. Temperature dependent magnetic susceptibility ($\chi$) measurements of EuRbFe$_4$As$_4$ single crystal**. (a) Temperature dependent $\chi$ measurements for Zero field cooled (ZFC) (solid blue circles) and field cooled (FC) (open blue circles) procedures over temperature range from above superconducting transition ($T_c$) to lower than magnetism transition ($T_m$). (b-d) The zoom-in of $\chi(T)$ plots in **a** focusing on the temperature close to $T_m$ for FC, ZFC and temperature close to $T_c$, respectively. The magnetic field applied was perpendicular to the FeAs plane and kept at 10 Oe during the measurement. Volume of the sample was measured to be 1.09×10$^{-4}$ cm$^3$.

**Extended data Fig. 2. Comparison of superconducting gap modulation below and above $T_m$**. **a,b** Topograph $T(\mathbf{r})$ and its associated Fourier transforms (FT). **c,d** Superconducting gap ($\Delta(\mathbf{r})$) map obtained over identical area as **a** and its associated FT, respectively. The black double-head arrows in **c** indicate the PDW stripe pattern. **e** Line cut of differential conductance (d$I$/d$V$) spectra along the dashed line as denoted in **a,c**. Inset is the d$I$/d$V$ intensities focusing on the coherence peak. **g-k** Corresponding results obtained over identical area at 16K, which is above $T_m$, with exactly the same setup conditions except the $T$. **f,l** Histograms of the gap size distributions from **c** and **i**, respectively. The gray curves represent the applied Gaussian fittings with $\Delta_{fit}$(4.5K) = 4.97meV, $\Delta_{fit}$(16K) = 4.91meV. Setup conditions: Topographs: $V_{set}$ = 12 mV, $I_{set}$ = 500 pA; Gap maps: $V_{set}$ = 12 mV, $I_{set}$ = 500 pA, $V_{exc}$ = 1 mV.

**Extended data Fig. 3. Additional large field-of-view $N(\mathbf{r}, E\sim\Delta_0)$ maps on different sample surfaces and regions.** Spatial resolutions of the $N(\mathbf{r}, E\sim\Delta_0)$ maps are (in unit of pixels/nm): top row from left to right: 1.37, 1.60, 2.06 and 1.80, respectively; bottom row from left to right: 1.37, 1.32, 1.37 and 1.30, respectively. The spatial resolution required to discern the modulations with an $8a_{Fe}$ periodicity in the $N(\mathbf{r})$ maps should be at least 0.93 pixels/nm. Dashed circles are vortices introduced by external magnetic fields. The last two rows are spatially averaged differential conductance (d$I$/d$V$) spectra taken over the corresponding regions as described in the first two rows. The blue dashed lines indicate the coherence peak positions. Setup conditions: $N(\mathbf{r}, E\sim\Delta_0)$ maps: Sample I, region A: $V_{set}$ = 12 mV, $I_{set}$ = 500 pA, $V_{exc}$ = 0.3 mV, 0T; Sample I, region B: $V_{set}$ = 6 mV, $I_{set}$ = 60 pA, $V_{exc}$ = 1 mV, 0T; Sample II, region A: $V_{set}$ = 5 mV, $I_{set}$ = 50 pA, $V_{exc}$ = 1 mV, 0T; Sample II, region B: $V_{set}$ = 5 mV, $I_{set}$ = 50 pA, $V_{exc}$ = 1 mV, 0T; Sample II, region C: $V_{set}$ = 5 mV, $I_{set}$ = 50 pA, $V_{exc}$ = 1 mV, 0.1T; Sample II, region D: $V_{set}$ = 5 mV, $I_{set}$ = 50 pA, $V_{exc}$ = 1 mV, 0.2T; Sample II, region E: $V_{set}$ = 5 mV, $I_{set}$ = 50 pA, $V_{exc}$ = 1 mV, 0T; Sample II, region F: $V_{set}$ = 5 mV, $I_{set}$ = 30 pA, $V_{exc}$ = 1 mV, 0.1T. Spectra: Sample I, region A: $V_{set}$ = 12 mV, $I_{set}$ = 500 pA, $V_{exc}$ = 0.3 mV, 0T; Sample I, region B:

$V_{set}$ = 20 mV, $I_{set}$ = 400 pA, $V_{exc}$ = 0.2 mV, 0T; Sample II, region A: $V_{set}$ = 30 mV, $I_{set}$ = 600 pA, $V_{exc}$ = 0.6 mV, 0T; Sample II, region B: $V_{set}$ = 12 mV, $I_{set}$ = 500 pA, $V_{exc}$ = 0.3 mV, 0T; Sample II, region C: $V_{set}$ = 15 mV, $I_{set}$ = 350 pA, $V_{exc}$ = 0.2 mV, 0.1T; Sample II, region D: $V_{set}$ = 30 mV, $I_{set}$ = 1500 pA, $V_{exc}$ = 0.15 mV, 0.2T; Sample II, region E $V_{set}$ = 30 mV, $I_{set}$ = 1500 pA, $V_{exc}$ = 0.15 mV, 0T; Sample II, region F: $V_{set}$ = 30 mV, $I_{set}$ = 1500 pA, $V_{exc}$ = 0.2 mV, 0.1T.

**Extended data Fig. 4. Large field-of-view $T$(r) maps and their Fourier transforms on different sample surfaces and regions acquired at different setup conditions.** Spatial resolutions of the $T(\mathbf{r})$ maps are (in unit of pixels/nm): first column from top to bottom: 2.26, 1.60, 1.80 and 1.68, respectively; third column from top to bottom: 2.06, 2.41, 1.53 and 1.37, respectively. The red squares denote the area where the corresponding $N(\mathbf{r}, E \sim \Delta_0)$ maps are taken as described in Figure S3. The second and fourth columns represent their associated Fourier transforms of the topographs. Area near $Q_{PDW}$ (black squares) are zoomed in for the visual purpose, and the green crosses locate the (¼, ¼) $2\pi/a_{Fe}$. The spatial resolution required to discern the modulations with an $8a_{Fe}$ periodicity in the $N(r)$ maps should be at least 0.93 pixels/nm. Setup conditions: Sample I, region A: $V_{set}$ = 12 mV, $I_{set}$ = 500 pA, $V_{exc}$ = 0.3 mV, 0T; Sample I, region B: $V_{set}$ = 6 mV, $I_{set}$ = 60 pA, $V_{exc}$ = 1 mV, 0T; Sample II, region A: $V_{set}$ = 5 mV, $I_{set}$ = 50 pA, $V_{exc}$ = 1 mV, 0T; Sample II, region B(column 1): $V_{set}$ = 50 mV, $I_{set}$ = 50 pA, 0T; Sample II, region B(column 3): $V_{set}$ = 5 mV, $I_{set}$ = 50 pA, 0T; region C: $V_{set}$ = 205 mV, $I_{set}$ = 50 pA, 0T; Sample II, region E(top): $V_{set}$ = 100 mV, $I_{set}$ = 40 pA, 0T; Sample II, region E(bottom): $V_{set}$ = 5 mV, $I_{set}$ = 30 pA, $V_{exc}$ = 1 mV, 0.1T.

**Extended data Fig. 5. Comparison of Fourier transforms of $N$(r,$E$) images below and above the magnetic transition temperature on sample II, region E.** The purple squares highlight the layers where strong $\mathbf{Q_{PDW}}$ peaks can be detected. Setup conditions: $V_{set}$ = 12 mV, $I_{set}$ = 500 pA, $V_{exc}$ = 0.3 mV, 0T.

**Extended data Fig. 6.** Raman spectroscopy measurements on Eu-1144. The left panel shows Raman spectra at different temperatures with offsets for clarity. The right panel shows the wavenumbers of the observed modes as a function of temperature. No mode softening is seen around the magnetic transition temperature of 16 K, and no additional modes are observed below the magnetic transition temperature.

**Extended data Fig. 7.** $C_4$ **rotational symmetry breaking below $T_m$. a,b,** Fourier transforms of $N(\mathbf{r}, E = 3.12$ meV) maps at 16 K (above $T_m$) and 4.5 K (below $T_m$), respectively, acquired over identical regions on the sample 2 (same sample as in Figs. 3 **j,k**). The quasiparticle interference (QPI) pattern at 4.5 K is obviously $C_2$, while that at 16 K is almost $C_4$. Both images have been 2-fold-symmetrized with respect to the $\mathbf{Q}_y$ axis. STM setup condition: $V_{set}$ = 12 mV, $I_{set}$ = 500 pA, $V_{exc}$ = 0.3 mV.

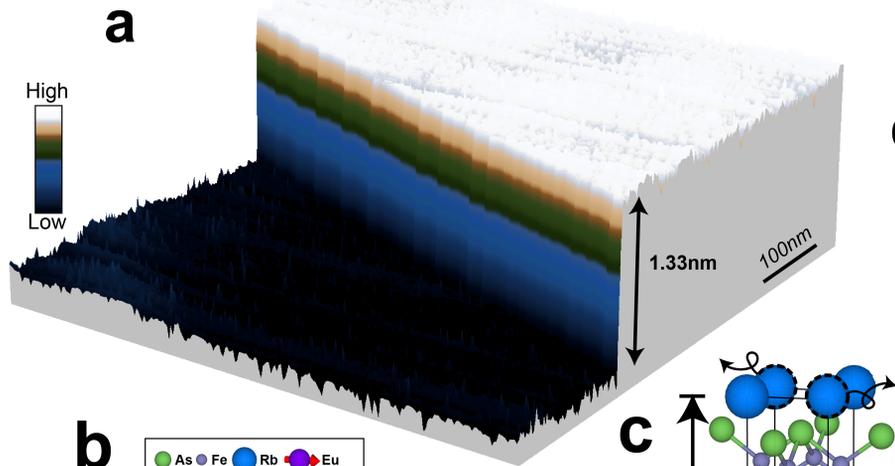
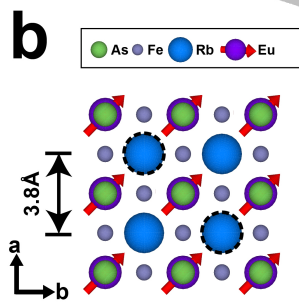
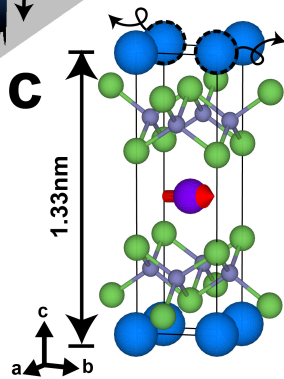
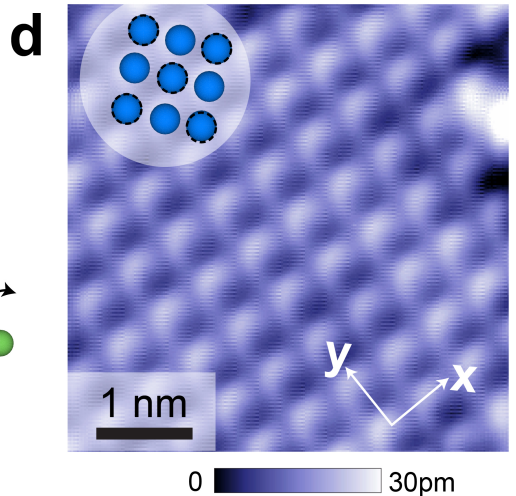
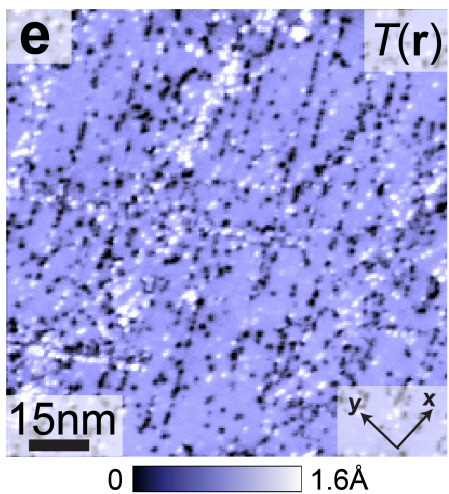
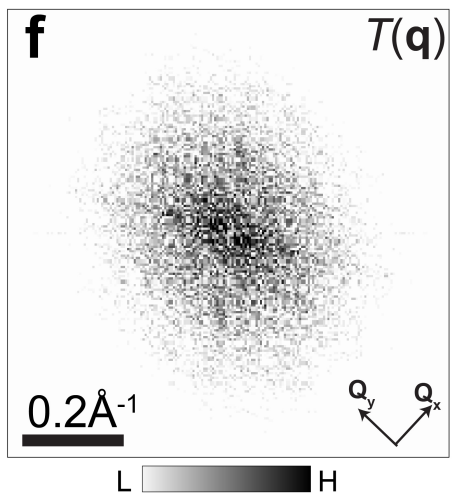
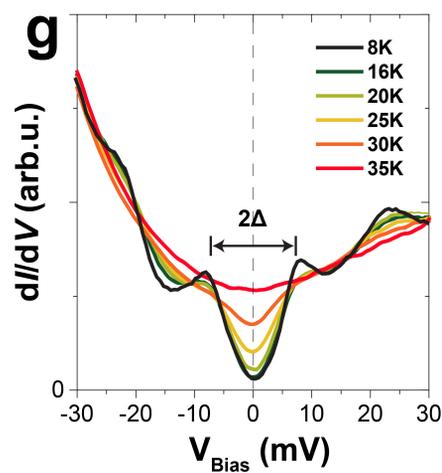

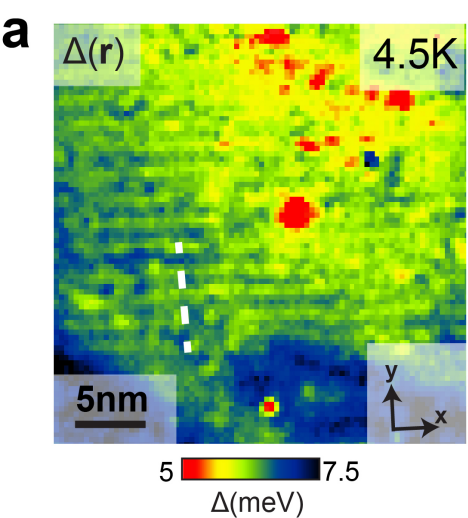 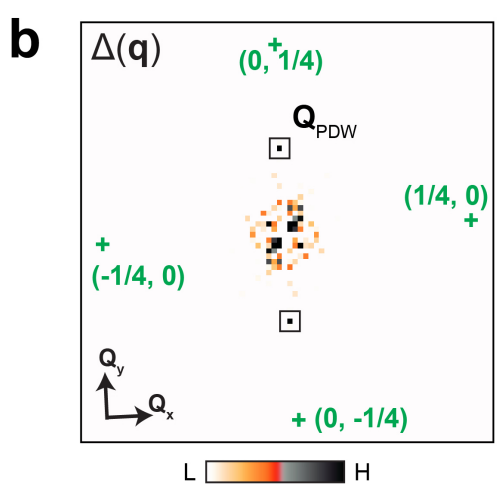 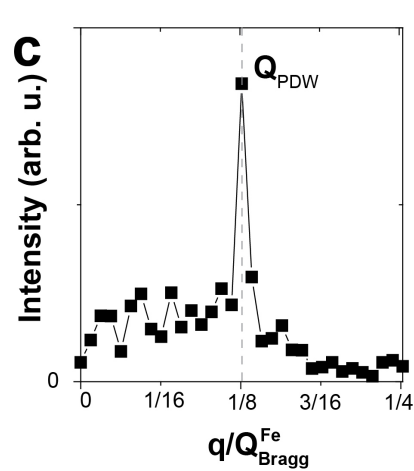
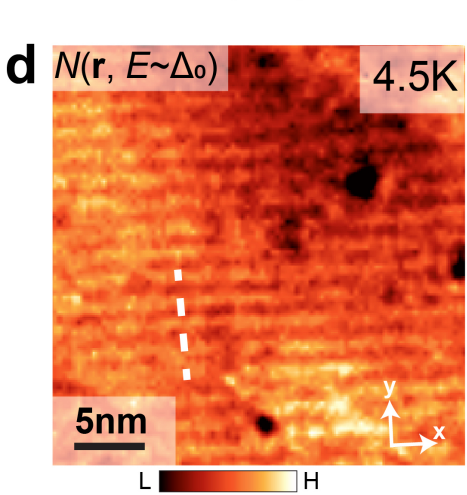 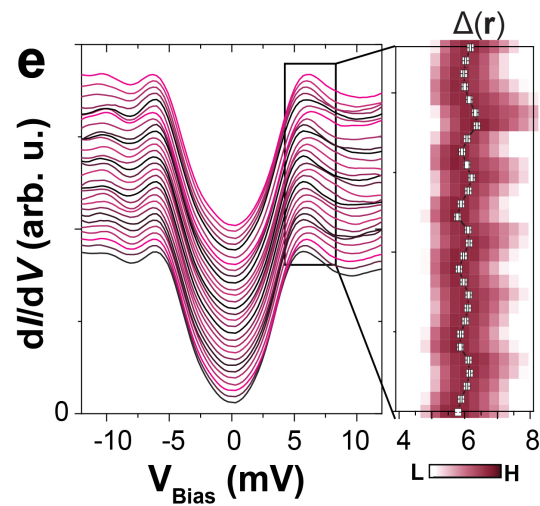 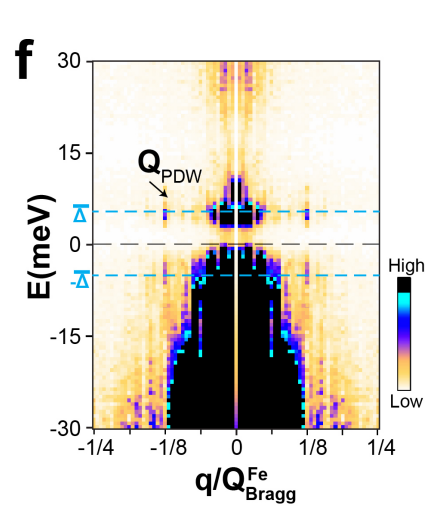

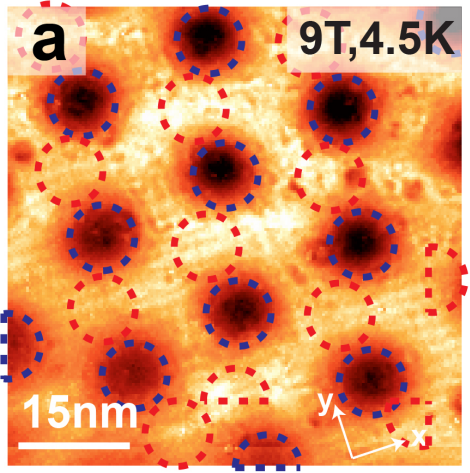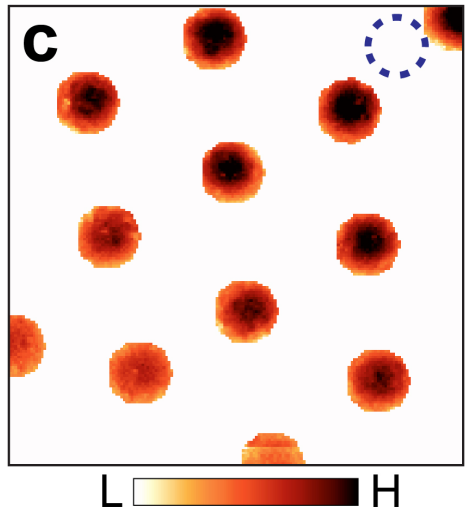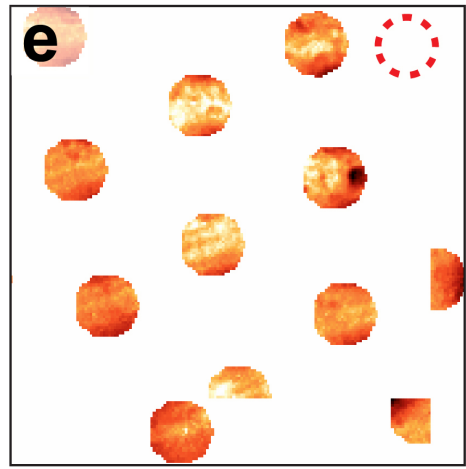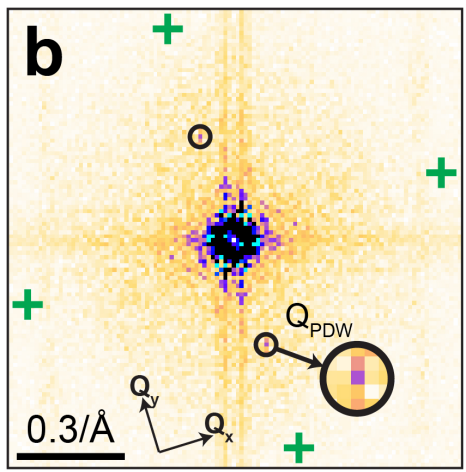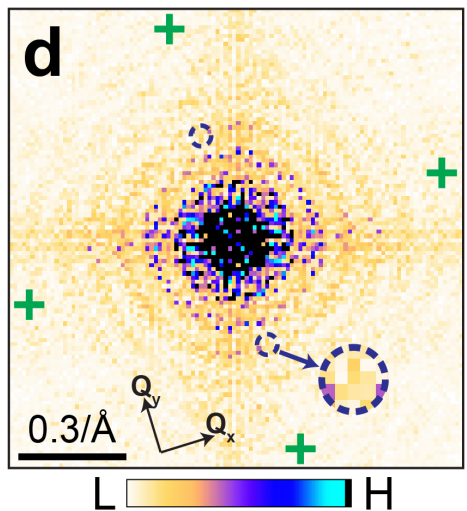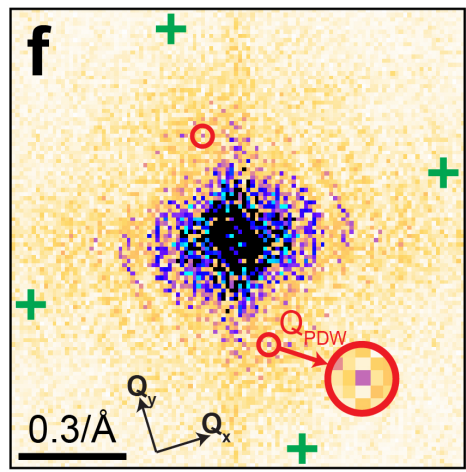

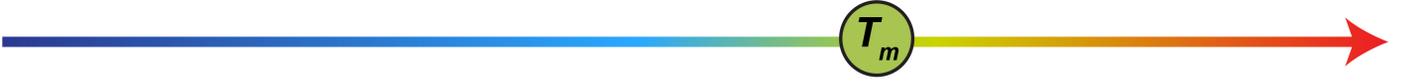
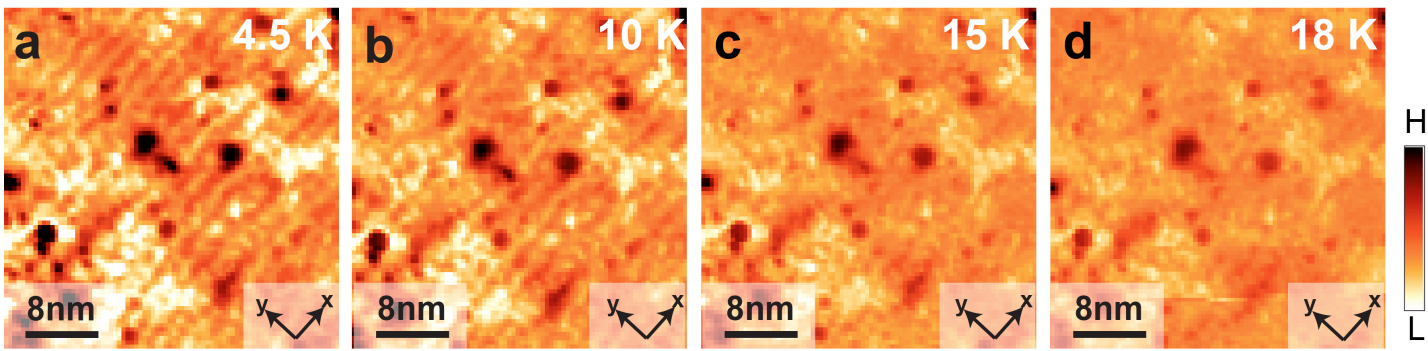
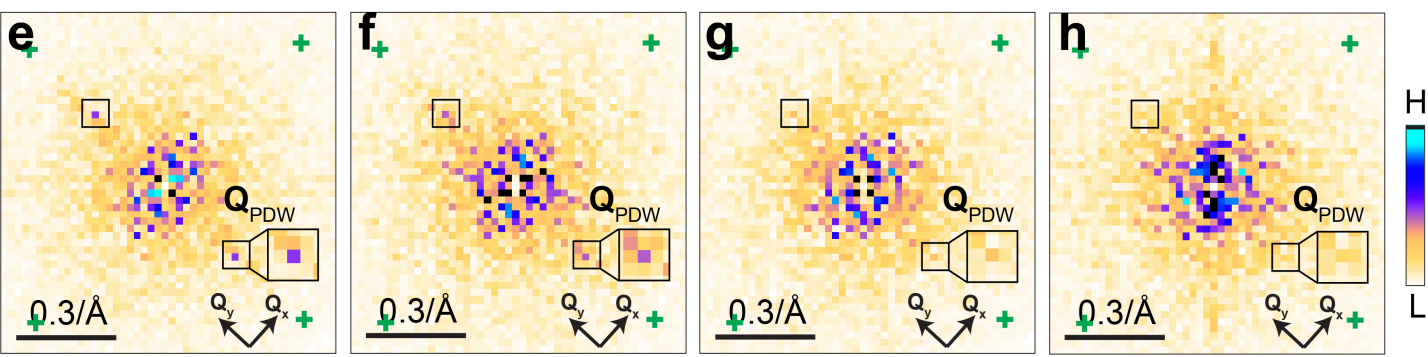
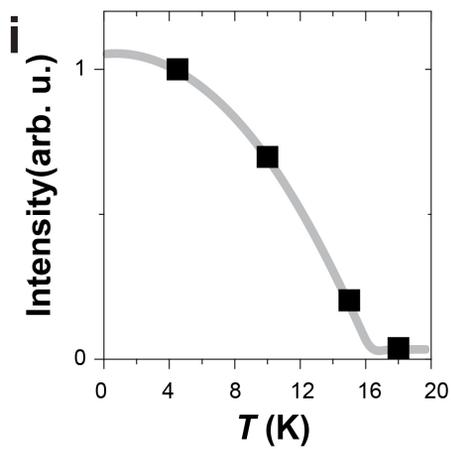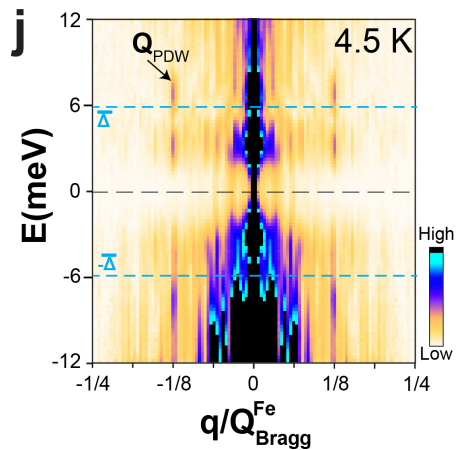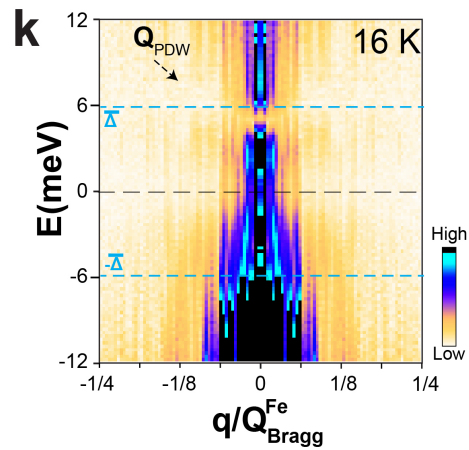